\documentclass[journal]{IEEEtran}
\pdfoutput=1
\usepackage{ifpdf}
\usepackage{subfigure}
\usepackage{graphicx}
\DeclareGraphicsExtensions{.pdf,.jpg,.png}
\usepackage[cmex10]{amsmath}
\usepackage{amssymb}
\usepackage{array}
\usepackage{mdwmath}
\usepackage{mdwtab}
\usepackage{pslatex}
\usepackage{xcolor}
\usepackage{textcomp}
\usepackage{placeins}
\usepackage{eqparbox}
\usepackage{url}
\usepackage{stfloats}
\usepackage{multicol}
\usepackage{subfigure}
\usepackage{float}
\usepackage{algorithmicx}
\usepackage{algpseudocode}
\usepackage{amsthm}
\usepackage{comment}
\usepackage{cite}
\usepackage{color, soul}
\usepackage{enumerate}
\usepackage{bbm}
\usepackage{threeparttable}
\usepackage{booktabs}
\usepackage{array}
\usepackage{balance}
\usepackage{color}
\usepackage{gensymb}

\begin{document}

\title{\vspace{-4mm}Capacity and Outage of Terahertz Communications with User Micro-mobility and Beam Misalignment\vspace{-1mm}}

\author{Vitaly Petrov, Dmitri Moltchanov, Yevgeni Koucheryavy, and Josep M. Jornet\vspace{-4mm}
\thanks{V. Petrov, D. Moltchanov, and Y. Koucheryavy are with Tampere University, Finland. J. M. Jornet is with Northeastern University, Boston, MA, USA.}}

\maketitle

\begin{abstract}
User equipment mobility is one of the primary challenges for the design of reliable and efficient wireless links over millimeter-wave and terahertz bands. These high-rate communication systems use directional antennas and therefore have to constantly maintain alignment between transmitter and receiver beams. For terahertz links, envisioned to employ radiation patterns of no more than few degrees wide, not only the macro-scale user mobility (human walking, car driving, etc.) but also the micro-scale mobility -- spontaneous shakes and rotations of the device -- becomes a severe issue. In this paper, we propose a mathematical framework for the first-order analysis of the effects caused by micro-mobility on the capacity and outage in terahertz communications. The performance of terahertz communications is compared with and without micro-mobility illustrating the difference of up to $1$\,Tbit/s or $75\%$. In response to this gap, it is finally shown how the negative effects of the micro-mobility can be partially addressed by a proper adjustment of the terahertz antenna arrays and the period of beam realignment procedure.
\end{abstract}\vspace{-0mm}
	

\vspace{-2mm}
\section{Introduction}\label{sect:01}\label{sec:intro}
\vspace{-1mm}


While the commercial fifth-generation (5G) mobile networks exploiting millimeter wave (mmWave) band are expected to appear within the next few years~\cite{mmWave_survey}, the researchers started targeting Terahertz (THz, $300$\,GHz--$3$\,THz) band communications as a technology enabler for novel 6G and beyond applications~\cite{akyildiz2014terahertz,rangan_6g_usecases_arxiv}. The sub-millimeter wavelength of THz frequencies promises ultra-large antenna arrays featuring thousands of elements at both transmitter and receiver sides~\cite{akyildiz2016realizing}. These arrays will be capable of creating extremely directional steerable antenna radiation patterns with the beamwidth of just a few degrees or even less. The later feature allows to overcome severe path loss and atmospheric absorption at THz frequencies and maintain adequate levels of signal-to-noise ratio (SNR) at the distances of up to a few tens of meters~\cite{jornet2011channel}.

On the negative side, the use of very narrow THz radiation patterns in mobile communications challenges the accuracy and speed of the employed beam steering procedures to follow the nodes mobility. For handheld and wearable THz devices, not only the conventional macro-mobility (human walking, car driving, etc.) but also the much less predictable micro-mobility (small-scale shakes and rotations) have to be considered~\cite{hur_dynamic_mmWave,petrov_nanocom}. These fast displacements of the mobile THz user equipment (THz-UE) may cause frequent misalignments of the highly-directional THz beams, consequently, leading to possible outages and degradation of the link capacity~\cite{peng_dynamic_thz}.


The topic has been partially studied to date. The capacity of a THz link with a fixed beam misalignment was evaluated in~\cite{priebe_fixed_thz}. Then, Priebe et al.~\cite{priebe_random_thz} performed ray-based simulations to explore the impact of typical human mobility on indoor THz systems. Approaches to analytically model the effect have been proposed, among others, in~\cite{wildman_random_mmWave} and~\cite{heath_random_mmWave}. Both works study a snapshot of the network, where the current misalignment is modeled as a random variable.
An indoor network with a THz access point (THz-AP) and walking human users has been simulated in~\cite{peng_dynamic_thz} and~\cite{carnegie_arxiv}. \textcolor{black}{An enhanced solution to maintain the mmWave beam alignment in the presence of micro-mobility has been presented in~\cite{revision_1}. The joint beam alignment and rate control algorithm for mmWave networks has been delivered in~\cite{revision_2}.} Finally, Hur et al.~\cite{hur_dynamic_mmWave} modeled the mmWave backhaul links with 2D mobility caused by the wind, however, no rotations have been considered.

To the best of the authors' knowledge, no mathematical framework quantifying the effect of the THz-UE micro-mobility (both displacements and rotations) on the mobile THz communications has been reported to date. We address this gap in the present article. The main contributions of this work are:
\begin{itemize}
\item To estimate the achieved capacity and fraction of time in outage of a THz link as a function of THz-UE micro-mobility parameters and antenna array characteristics, a unified mathematical framework is proposed based on the probability and random walk theories.
\item To numerically characterize the system performance when accounting for the THz-UE micro-mobility, a mathematical methodology is employed. \textcolor{black}{It is particularly shown that the capacity of the THz link can decrease by up to $75\%$ in the presence of THz-UE micro-mobility. It is also demonstrated that a proper adjustment of the antenna array sizes using our framework results in up to $500$\,Gbit/s capacity improvement.}
\end{itemize}

\vspace{-2mm}
\section{System Model}\label{sect:02}
\label{sec:system_model}
\vspace{-1mm}

\subsubsection{Deployment and Propagation Model}

\begin{figure}[!t]
  \centering
  \vspace{-0mm}
  \includegraphics[width=0.8\columnwidth]{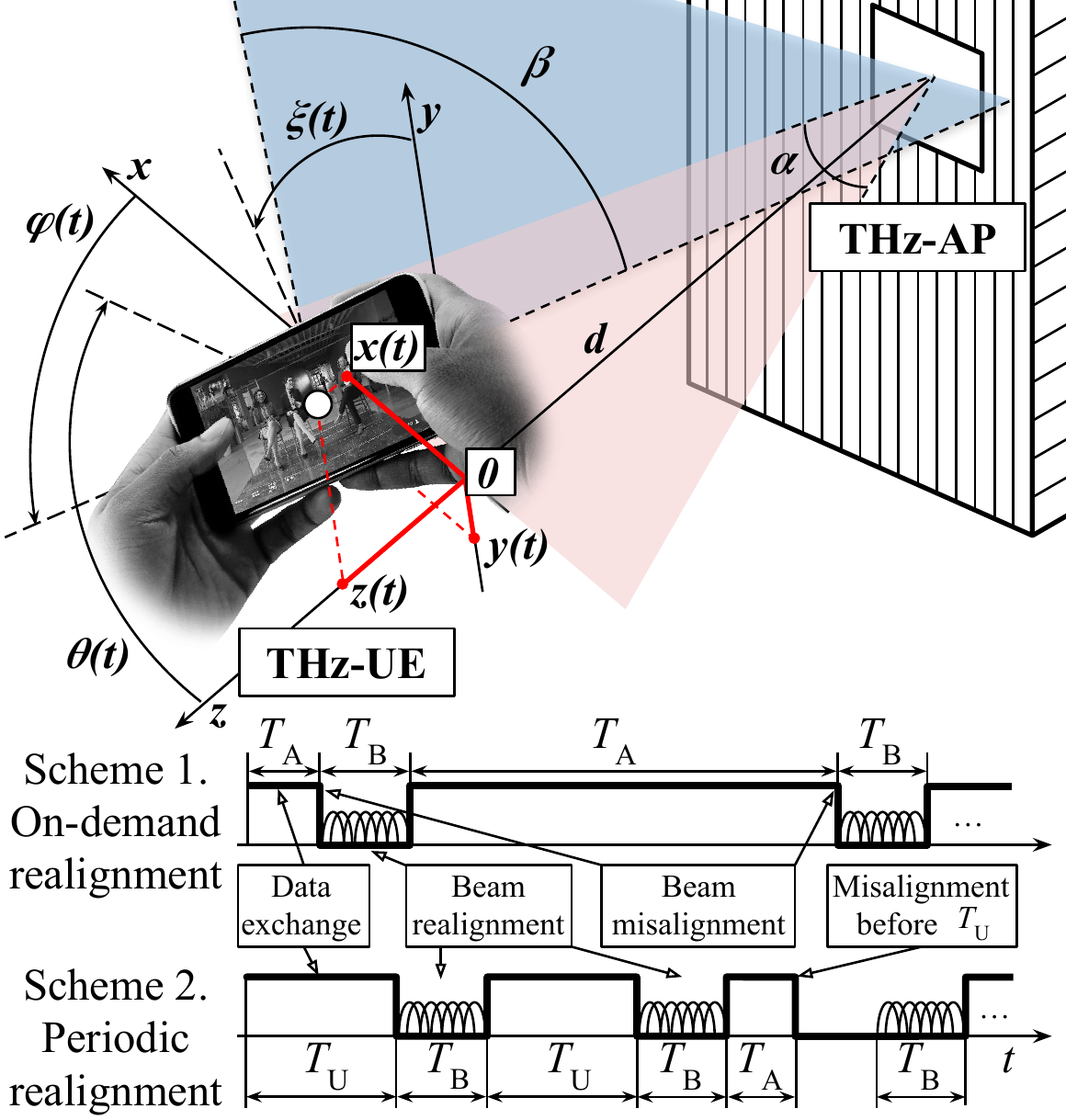}
  \vspace{-3mm}
  \caption{Modeled THz communications with micro-mobility of THz-UE.}
  \vspace{-3mm}
  \label{fig:scenario}
\end{figure}

We consider a single link between a stationary THz-AP and a handheld THz-UE separated by $d$ meters (see Fig.~\ref{fig:scenario}). To model the narrow THz beams, we assume that both THz-AP and THz-UE are equipped with 2D antenna arrays of $N_{\text{A}}$$\times$$N_{\text{A}}$ and $N_{\text{U}}$$\times$$N_{\text{U}}$ ($N_{\text{A}} \gg N_{\text{U}}$) elements generating 3D sectored antenna radiation patterns with the gains of $G_{\text{A}} = N_{\text{A}}^2$, $G_{\text{U}} = N_{\text{U}}^2$. \textcolor{black}{The corresponding antenna radiation pattern angles for THz-AP and THz-UE are approximated as $\alpha = \frac{102\pi}{180N_{\text{A}}}$ and $\beta = \frac{102\pi}{180N_{\text{U}}}$~\cite{balanis2016antenna}}

We employ a THz-specific propagation model from~\cite{jornet2011channel}. According to it, the SNR at the THz-UE, $S_{\text{U}}$, is written as
\begin{align}\label{eqn:snr}
\hspace{-2mm}S_{\text{U}} = P_{\text{A}}G_{\text{A}}G_{\text{U}}\frac{c^2}{16 \pi^2 f^2 N_{0}(B)}d^{-2}e^{-K d},
\end{align}
where $f$ is the carrier frequency, $B$ is the bandwidth, $K$ is an absorption coefficient for the band, $N_{0}(B)$ is the noise level.

\subsubsection{Micro-mobility Model}

The micro-mobility of the THz-UE is modeled as a combination of random displacements processes over OX and OY axis in Fig.~\ref{fig:scenario}, $x(t)$ and $y(t)$, together with the random rotation processes $\phi(t)$ and $\theta(t)$. These stochastic processes are modeled as independent random walks and parameterized with the mean displacement after one second of action: $\Delta{}x$, $\Delta{}y$, $\Delta{}\phi$, and $\Delta{}\theta$, respectively.

To simplify the model, we do not account for any micro-mobility along the OZ axis in Fig.~\ref{fig:scenario}, $z(t)$, or device rotations over this axis, $\xi(t)$. The reason is that for the considered scenario neither cm-scale movements along the communication path nor any rotations perpendicular to the communication link have any notable effect on the link-level performance.


\subsubsection{Beam Alignment Schemes}\label{sec:schemes}

The design of efficient beam alignment mechanisms for THz communications is still in progress, so our mathematical framework is independent of the choice of the specific beam alignment procedure. We only assume that its duration is constant and equals to $T_{\text{B}}$.

In practice, $T_{\text{B}}$ depends on many factors, including the employed algorithms, the sizes of the antenna arrays, and the time it takes to assess a single configuration of beams. \textcolor{black}{To account for these factors in our numerical study, we model a sequential beam alignment (similar to IEEE 802.11ad \cite{ieee_802_11ad}) with the complexity of $T_\text{B} = (N_{\text{A}}^2 + N_{\text{U}}^2) \delta$, where $\delta$, termed further as a beam steering delay, is the time duration it takes the THz node to scan one direction plus the time to change the direction of the THz beam. One can use different formulas for $T_{\text{B}}$ to model other beam alignment procedures.}

Two schemes are analyzed in our study (see Fig.~\ref{fig:scenario}):
\begin{itemize}
\item \emph{Scheme 1. On-demand alignment}. Beam alignment procedure runs every time the THz-UE micro-mobility results in an outage. This scheme reflects the design of WLANs.
\item \emph{Scheme 2. Periodic alignment}. Beam realignment procedure runs periodically with a period $T_{\text{U}} + T_{\text{B}}$. This scheme reflects the cellular-style systems with centralized control.
\end{itemize}

\textcolor{black}{In the following section, we present our mathematical framework to analyze and compare the characteristics of these schemes in the presence of THz-UE micro-mobility. For simplicity, we assume that the detection of outage (beam misalignment) is instantaneous. However, the framework can be further extended to account for non-zero time to detect an outage by introducing an additional component in $T_{\text{B}}$.}

\vspace{-1mm}
\section{The Proposed Approach}\label{sect:03}\label{sec:analysis}

\textcolor{black}{In this section, we detail our mathematical framework to analyze and compare the characteristics of Scheme 1 and Scheme 2 in the presence of THz-UE micro-mobility. We first derive the probability density function (pdf) for the amount of time for micro-mobility to cause beam misalignment in Subsection~\ref{sec:fpt}. Later, in Subsection~\ref{sec:schemes_analysis}, we apply the obtained pdf to evaluate the performance of the THz link.}

\vspace{-3mm}
\subsection{Beam Misalignment due to Micro-Mobility}\label{sec:fpt}
\vspace{-1mm}

\subsubsection{Micro-mobility components} 

Assume that at the time $t=0$ THz-AP and THz-UE beams are perfectly aligned. To represent each mobility component, $\phi(t)$, $\theta(t)$, $x(t)$ $y(t)$, we use the limiting behavior of the random walk -- a Brownian motion process described by the following equation~\cite{redner2001guide}
\begin{align}\label{eqn:conv_diff}
\frac{\partial{}c(x,t)}{\partial{}t}-D\frac{\partial^2c(x,t)}{\partial{}x}=0,\quad{}t>0,
\end{align}
where the ``concentration'' $c(x,t)$ is interpreted as the probability of having a diffusing point at coordinate $x$ at time $t$, $D=(\Delta{}x)^2/2\Delta{}t$ is the diffusion coefficient, $\Delta{x}$ is the displacement corresponding to time increment $\Delta{t}$. Note that mobility processes are associated with diffusion coefficients $D_\phi$, $D_\theta$, $D_x$, $D_y$. Considering second as a unit time, we define $D_x=(\Delta{x})^2/2$, where $\Delta{x}$ is the displacement per second. The same applies for the rest of the processes.


\subsubsection{Micro-mobility $x(t)$ and $y(t)$}

We first characterize the effect of micro-mobility induced by $x(t)$ and $y(t)$. As $N_{\text{A}} \gg N_{\text{U}}$, the THz-AP beam is narrower than the one from THz-UE. So, the misalignment will happen when either $x(t)$ or $y(t)$ leaves the THz-AP beam with the geometrical boundaries $[-M_{\text{XY}},+M_{\text{XY}}]$ where $M_{\text{XY}} = d\tan(\alpha/2) = d \tan(\frac{102\pi}{360N_{\text{A}}})$~\cite{balanis2016antenna}.

Consider first $x(t)$. The pdf of the first passage time (FPT) to the symmetric boundaries $\pm{}M_{\text{XY}}$ is given by~\cite{redner2001guide}
\begin{align}\label{eqn:fptSingle}
f_{X}(t)&= \sum^{\infty}_{n = -\infty}\frac{M_{\text{XY}}[4n+1] \left[ \exp \left(-\frac{M_{\text{XY}}^2}{4 D_x  t}\right) \right]^{(4n+1)^2}}{\sqrt{\pi  {D_x}  t^3}},  \,t>0.
\end{align}


Integrating (\ref{eqn:fptSingle}), the cumulative distribution function (cdf) is
\begin{align}\label{eqn:fptSingleCDF}
F_X(t)=\sum_{n=-\infty}^{\infty}\frac{(4 n+1) \left[\text{erf}\left(\frac{M_{\text{XY}} \sqrt{\frac{8 n^2+4 n+1}{D_x t}}}{\sqrt{2}}\right)-1\right]}{\sqrt{4 n^2+2 n+1/2}},
\end{align}
where $\text{erf}(\cdot)$ is the error function.

{
\color{black}
Similarly to (\ref{eqn:fptSingle}) we define pdf of the FPT for $y(t)$, $f_Y(t)$. Now, the time it takes for a connection to be lost due to either $x(t)$ or $y(t)$ is the minimum of the considered two random variables (RV). Hence, the pdf of this minimum, $f_{T_{X,Y}}(t)$, is
\begin{align}\label{eqn:minpdfXY}
f_{T_{X,Y}}(t)&=f_{X}(t)[1-F_{Y}(t)]+f_{Y}(t)[1-F_{X}(t)].
\end{align}
}

\subsubsection{Micro-mobility $\phi(t)$ and $\theta(t)$}

We now include angular mobility in our analysis. The misalignment caused by rotations happens when THz-AP appears outside of the turning THz-UE beam. In general, the particular angle that THz-UE has to turn over $\phi$ and $\theta$ to cause misalignment depends on the current location of the THz-UE, $x(t)$ and $y(t)$. At the same time, for any current $\Delta{}x$ and $\Delta{}y$, the bounding angle $M_{\phi \theta}$ is $M_{\phi \theta} \in [\frac{\beta}{2}-\frac{\alpha}{2}, \frac{\beta}{2}+\frac{\alpha}{2}]$, $\alpha = \frac{102\pi}{180N_{\text{A}}}$, $\beta = \frac{102\pi}{180N_{\text{U}}}$ (see Fig.~\ref{fig:scenario}).

As $N_{\text{A}} \gg N_{\text{U}}$, the first term dominates, and the range is relatively small. For clarity, we use an upper bound on $M_{\phi \theta}$, $M_{\phi \theta} = \frac{102\pi}{360}(1 / N_{\text{U}} + 1/N_{\text{A}})$, so when either $\phi(t)$ or $\theta(t)$ reaches $\pm M_{\phi \theta}$, the alignment is lost. We can, finally, characterize the FPT pdfs for $\phi(t)$ and $\theta(t)$ -- $f_{\Phi}(t)$ and $f_{\Theta}(t)$ -- using (\ref{eqn:fptSingle}), while $f_{T_{\Phi,\Theta}}(t)$ is obtained as a minimum of these two RVs.

\begin{figure*}[!t]
\vspace{-5mm}
\footnotesize
\setcounter{equation}{7}
\begin{align}\label{eqn:overallpdf}
\hspace{-3mm}f_{T_{\text{A}}}(t)=\frac{\frac{e^{-\frac{(\log (t)-\mu_x)^2}{2 \sigma_x^2}}}{\sigma_x} \left[2-\text{erfc}\left(\frac{\mu_y-\log (t)}{\sqrt{2} \sigma_y}\right)\right]+\frac{e^{-\frac{(\log (t)-\mu_y)^2}{2 \sigma_y^2}}}{\sigma_y} \left[2-\text{erfc}\left(\frac{\mu_x-\log (t)}{\sqrt{2} \sigma_x}\right)\right]}{ 2   \sqrt{2 \pi } t       \left[1-\frac{1}{2} \text{erfc}\left(\frac{\mu_\phi-\log (t)}{\sqrt{2} \sigma_\phi}\right)+\frac{1}{2} \text{erfc}\left(\frac{\mu_\theta-\log (t)}{\sqrt{2} \sigma_\theta}\right)\right]^{-1}}
+
\frac{\frac{e^{-\frac{(\log (t)-\mu_\phi)^2}{2 \sigma_\phi^2}}}{\sigma_\phi} \left[2-\text{erfc}\left(\frac{\mu_\theta-\log (t)}{\sqrt{2} \sigma_\theta}\right)\right]+\frac{e^{-\frac{(\log (t)-\mu_\theta)^2}{2 \sigma_\theta^2}}}{\sigma_\theta} \left[2-\text{erfc}\left(\frac{\mu_\phi-\log (t)}{\sqrt{2} \sigma_\phi}\right)\right]}{ 2   \sqrt{2 \pi } t       \left[1-\frac{1}{2} \text{erfc}\left(\frac{\mu_x-\log (t)}{\sqrt{2} \sigma_x}\right)+\frac{1}{2} \text{erfc}\left(\frac{\mu_y-\log (t)}{\sqrt{2} \sigma_y}\right)\right]^{-1}}
\end{align}
\normalsize
\hrulefill
\setcounter{equation}{5}
\vspace{-4mm}
\end{figure*}

\subsubsection{Aggregated micro-mobility}
Let us define $T_{\text{A}}$ as an RV representing the FPT for reaching any of the boundary conditions $\pm M_{\text{XY}}$ or $\pm M_{\Phi \Theta}$. Now, we characterize $f_{T_{\text{A}}}(t)$ -- the pdf of this FPT -- as the minimum of two RVs, i.e., 
\begin{align}\label{eqn:vit1}
f_{T_{\text{A}}}(t) = f_{T_{X,Y}}(t) [1-F_{T_{\Phi,\Theta}}(t)]+f_{T_{\Phi,\Theta}}(t)[1-F_{T_{X,Y}}(t)].
\end{align}

{
\color{black}
While the sought pdf can be precisely calculated, the resulting equation involves \emph{quadruple infinite sums} over $n$ resulting in high computational complexity. At the same time, despite the fact that the pdf $f_{X}(t)$ has a complex expression, the mean and variance of the FPT are finite and equal to $M^2_{\text{XY}}/2D_{x}$ and $M^4_{\text{XY}}/6D^2_{x}$, respectively~\cite{sherif1980first}. Therefore, recalling that the FPT to $\pm M_{\text{XY}}$ with individual micro-mobility component can be well approximated with Lognormal distribution featured by the same first moments~\cite{sherif1980first,margolin2004continuous}\footnote{\textcolor{black}{To check the proposed approximation, we have utilized the original distribution specified in~(\ref{eqn:fptSingle}) and the approximating Lognormal distribution to generate samples of equal size and then performed statistical chi-square test. The hypothesis $H_0$ was that both samples belong to the same distribution against an alternative hypothesis, $H_1$, that the distributions differ. With the level of significance set to $\alpha=0.05$ the test shows that for the ranges of input parameters considered in the paper samples drawn from original distribution and the approximating one belong to the same distribution.}}, with the use of (\ref{eqn:minpdfXY}) we have
}
\begin{align}\label{eqn:minXYlog}
&f_{T_{X,Y}}(t)=\frac{1}{{2 \sqrt{2 \pi } \sigma_y \sigma_x  t}}\Bigg[\sigma_x  e^{-\frac{(\mu_x-\log (t))^2}{2 \sigma_y^2}} \bigg[\text{erf}\left(\frac{\mu_x -\log(t)}{\sqrt{2} \sigma_x }\right)+1\bigg]+\nonumber\\
&+\sigma_y e^{-\frac{(\mu_x -\log (t))^2}{2 \sigma_x^2}} \bigg[\text{erf}\left(\frac{\mu_x-\log (t)}{\sqrt{2} \sigma_y}\right)+1\bigg]\Bigg],\quad{}t>0,
\end{align}
where $\mu_{x} = \log(\frac{2M_{\text{XY}}}{\Delta{}x}) - \frac{1}{2}\log(\frac{5}{3})$, $\mu_{y} = \log(\frac{2M_{\text{XY}}}{\Delta{}y}) - \frac{1}{2}\log(\frac{5}{3})$, $\sigma_x = \sigma_y=\sqrt{\log(5/3)}$, and also $\mu_{\phi}$$=$$\log(\frac{2M_{\text{XY}}}{\Delta{}\phi})$$-$$\frac{1}{2}\log(\frac{5}{3})$, $\mu_{\theta}=\log(\frac{2M_{\text{XY}}}{\Delta{}\theta})-\frac{1}{2}\log(\frac{5}{3})$, $\sigma_{\phi}=\sigma_{\theta}=\sqrt{\log(5/3)}$. 

{
\color{black}
Now, to determine the PFT for the aggregated mobility we apply (\ref{eqn:vit1}) to arrive at (\ref{eqn:overallpdf}), where $\text{erfc}(\cdot)$ is the complementary error function, while cdfs $F_{T_{X,Y}}(t)$ and $F_{T_{\Phi,\Theta}}(t)$ are obtained by integrating the corresponding pdfs.
}

\vspace{-4mm}
\subsection{THz Link Metrics with Micro-Mobility}\label{sec:schemes_analysis}
\vspace{-1mm}

\textcolor{black}{Once the random time the aggregated micro-mobility of THz-UE leads to beam misalignment, $T_{\text{A}}$, is characterized, we proceed with deriving the THz link metrics of interest. We particularly obtain: (i) the fraction of time THz-UE is in outage, including the time spent for the beam realignment; (ii) the mean spectral efficiency (SE) and capacity of the THz link; and (iii) the mean time to the first beam misalignment caused by the micro-mobility of THz-UE.}

\subsubsection{Scheme 1} Consider first the scheme, where the beam alignment procedure is only invoked when THz-UE gets to an outage condition. The fraction of link outage time is $p_{\text{O},1}=T_\text{B}/(T_\text{A}+T_\text{B})$, where $T_\text{B}$ is the duration of the beam alignment procedure. Using law of the unconscious statistician we have
\setcounter{equation}{7}
\begin{align}\label{eqn:out1}
p_{\text{O},1}=\int_{0}^{\infty}\frac{T_\text{B}}{t+T_\text{B}}f_{T_\text{A}}(t)dt.
\end{align}

\textcolor{black}{To obtain mean spectral efficiency (SE) and capacity of the THz link, $E[L_{1}]$ and $E[C_{1}]$\footnote{\textcolor{black}{E[X] stands for the mean value of X.}}}, we recall that both are zero for~$p_{O,1}$ and determined by $S_{\text{U}}$ from (\ref{eqn:snr}) for the rest of the time as
\begin{align}\label{eqn:cap}
L_{\max}=\log_{2}(1+S_U),\quad{}C_{\max}=B\log_{2}(1+S_U),
\end{align}
where $L_{\max}$ and $C_{\max}$ represent the SE and link capacity, respectively, when the beams are aligned.


The mean SE and capacity with \emph{Scheme~1} thus are
\begin{align}
E[L_1]=(1-p_{O,1})L_{\max},\quad{}E[C_1]=(1-p_{O,1})C_{\max}.
\end{align}

Finally, the mean time to the first beam misalignment is
\begin{align}
E[T_1]=\int_{0}^{\infty}f_{T_\text{A}}(t)dt.
\end{align}

\subsubsection{Scheme 2} We now study the second scheme, where beam alignment is invoked after a fixed interval $T_\text{U}$. Here, $T_{\text{O},2}$ -- the outage duration in a period $T_{\text{U}}+T_{\text{B}}$ -- depends on whether the time to misalignment is greater or smaller than~$T_\text{U}$:
\begin{align}\label{eqn:t02}
T_{\text{O},2}=
\begin{cases}
\frac{T_\text{B}}{T_\text{U}+T_\text{B}},&T_\text{A}\geq{}T_\text{U},\\
\frac{T_\text{U}+T_\text{B}-T_\text{A}}{T_\text{U}+T_\text{B}},&T_\text{A}<T_\text{U}.\\
\end{cases}
\end{align}

The probability $Pr\{T_\text{A}<T_\text{U}\}$ is directly obtained as
\begin{align}\label{eqn:q}
Pr\{T_\text{A}<T_\text{U}\} = \int_{0}^{T_\text{U}}f_{T_\text{A}}(t)dt = F_{T_\text{A}}(T_{\text{U}}).
\end{align}

{
\color{black}
The conditional pdf of the time to misalignment provided that it is less than $T_\text{U}$ is $f_{T_\text{A}}(t)/F_{T_\text{A}}(T_{\text{U}})$. Now, the fraction of time in outage, $p_{\text{O},2}$, can be obtained by weighting the means of two fractions in (\ref{eqn:t02}) as
\begin{align}
p_{\text{O},2}=\frac{T_\text{B}[1-F_{T_\text{A}}(T_{\text{U}})]}{T_\text{U}+T_\text{B}}+\int_{0}^{T_\text{U}}\frac{T_\text{U}+T_\text{B}-t}{T_\text{U}+T_\text{B}}f_{T_\text{A}}(t)dt.
\end{align}
}

The SE and capacity of the THz link with \emph{Scheme~2} are
\begin{align}
E[L_2]=(1-p_{\text{O},2})L_{\max},\,E[C_2]=(1-p_{\text{O},2})C_{\max},
\end{align}
where $L_{\max}$ and $C_{\max}$ are given in (\ref{eqn:cap}).

{
\color{black}
In contrast to \emph{Scheme 1}, the mean time to the first beam misalignment, $E[T_2]$ in this case can be longer than $T_\text{A}$. Observing that $T_2$ consist of several consecutive intervals of duration $T_\text{U}+T_\text{B}$ plus a fraction of the last interval where the actual misalignment will happen, we arrive at
}
\begin{align}
E[T_2] = \frac{1 - F_{T_\text{A}}(T_{\text{U}})}{F_{T_\text{A}}(T_{\text{U}})}\left( T_\text{U} + T_\text{B} \right) + \frac{1}{  F_{T_\text{A}} }\int_{0}^{T_\text{U}} t  f_{T_{\text{A}}}(t) dt.
\end{align}
\vspace{-4mm}
\section{Numerical Results}\label{sect:04}\label{sec:numerical_results}
\vspace{-1mm}

In this section, our derived results are numerically elaborated. Following the recent IEEE standard for wireless communications over low-THz band~\cite{ieee_802_15_3d}, we assume $f = 0.3$\,THz, $B = 50$\,GHz, and $P_{A} = 20$\,dBm. The THz beam steering delay, $\delta$, is set to $5$\,mcs. To better contrast the impact of Cartesian and angular micro-mobility, we assume $\Delta{}x=\Delta{}y$ and $\Delta{}\phi = \Delta{}\theta$. For illustrative purposes, we also take certain representative values for $\Delta{}x$ and $\Delta{}\phi$ from the measurement campaign reported in~\cite{petrov_nanocom}. Specifically, $\Delta{}x = 0.1$\,m and $\Delta{}\phi = 4\degree$ is typical for playing a dynamic game on a smartphone, while $\Delta{}x = 0.01$\,m; $\Delta{}\phi=3\degree$ correspond to the video watching scenario.

\begin{figure*}[!t]
\centering
\vspace{-7mm}
\subfigure[{Outage vs. $T_{\text{U}}$ for \emph{Scheme 2}}]
{
	\includegraphics[width=0.31\textwidth]{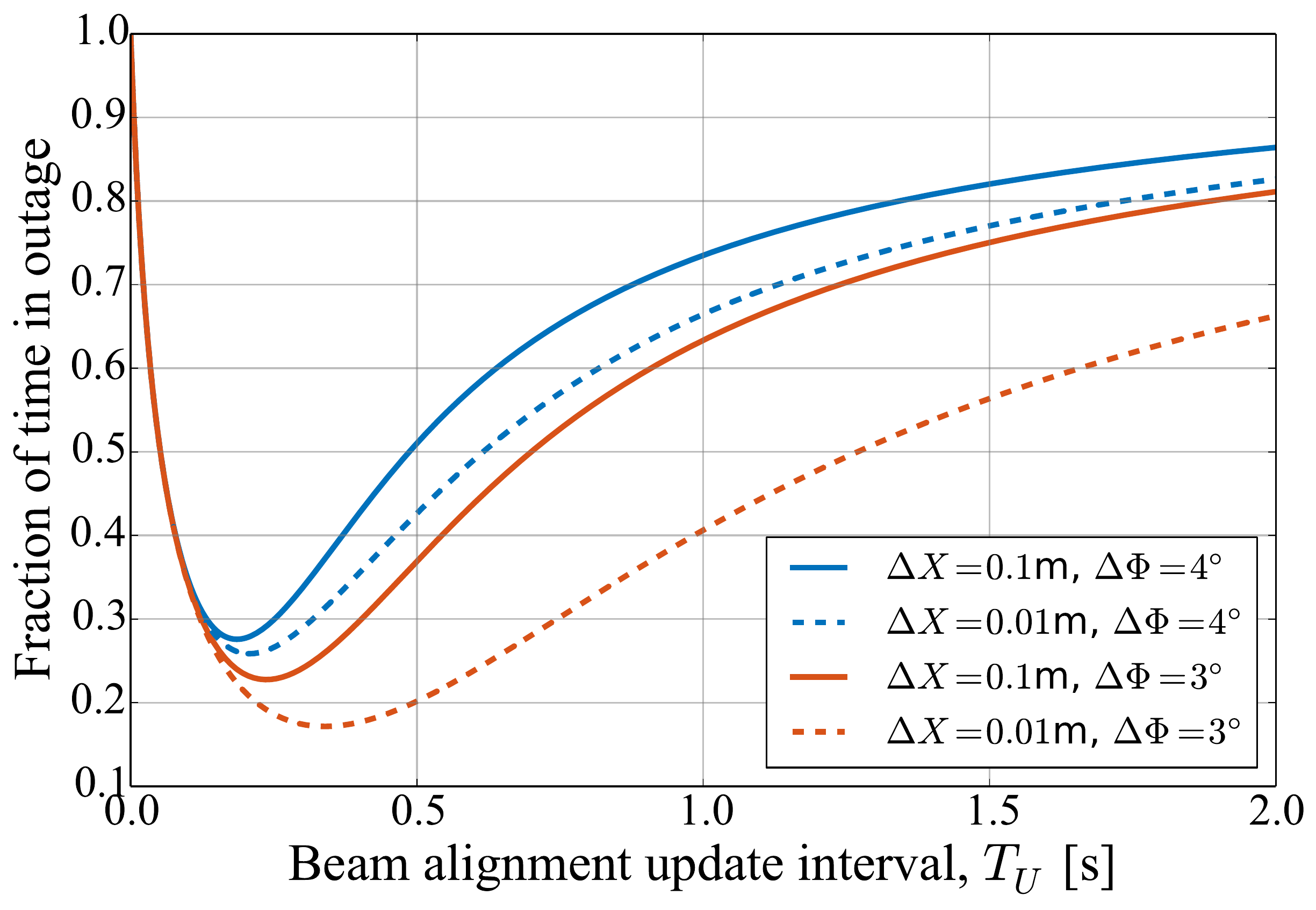}
	\label{fig:plot1}
}
\subfigure[{Mean SE vs. $N_{\text{U}}$ for both schemes}]
{
	\includegraphics[width=0.31\textwidth]{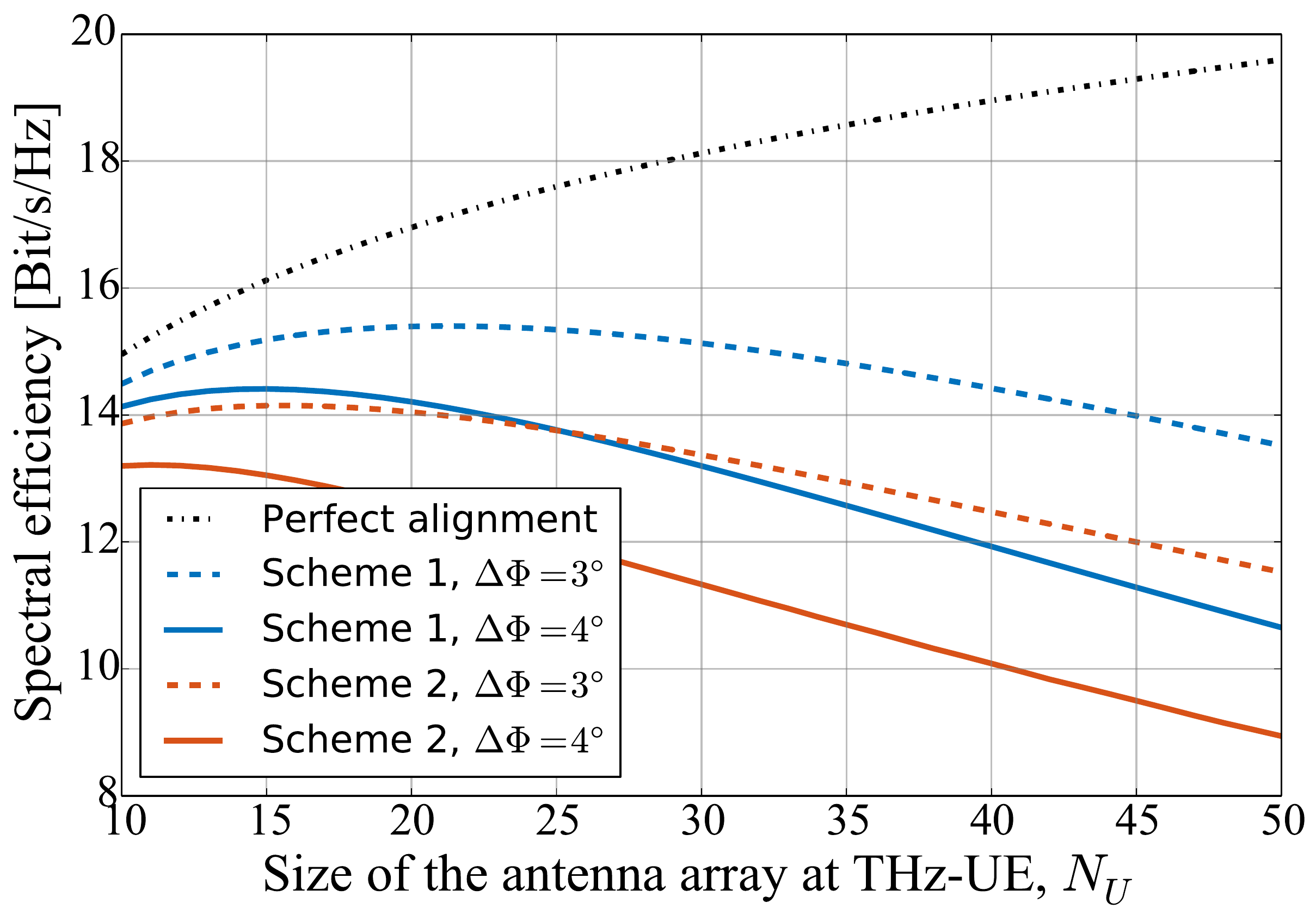}
	\label{fig:plot2}
}
\subfigure[{Mean SE vs. $N_{\text{A}}$ for both schemes}]
{
	\includegraphics[width=0.31\textwidth]{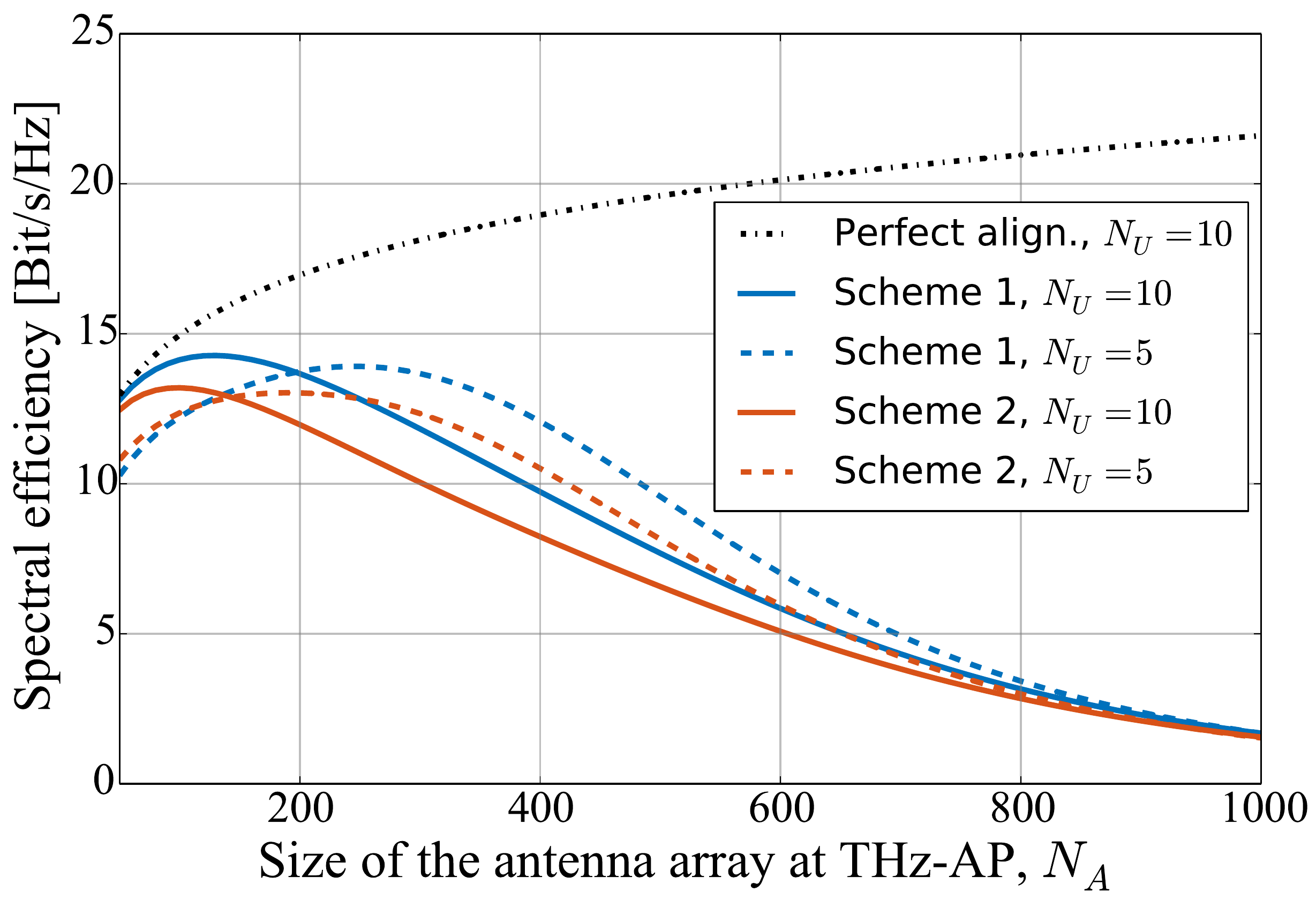}
	\label{fig:plot3}
}
\vspace{-2mm}
\caption{Adjustment of $T_{\text{U}}$, $N_{\text{U}}$, and $N_{\text{A}}$ for the considered levels of THz-UE micro-mobility.}
\label{fig:plots_optimizing}
\vspace{-4mm}
\end{figure*}

\begin{figure*}[!t]
\centering
\subfigure[{Outage vs. $\Delta{}\phi$ for both schemes}]
{
	\includegraphics[width=0.33\textwidth]{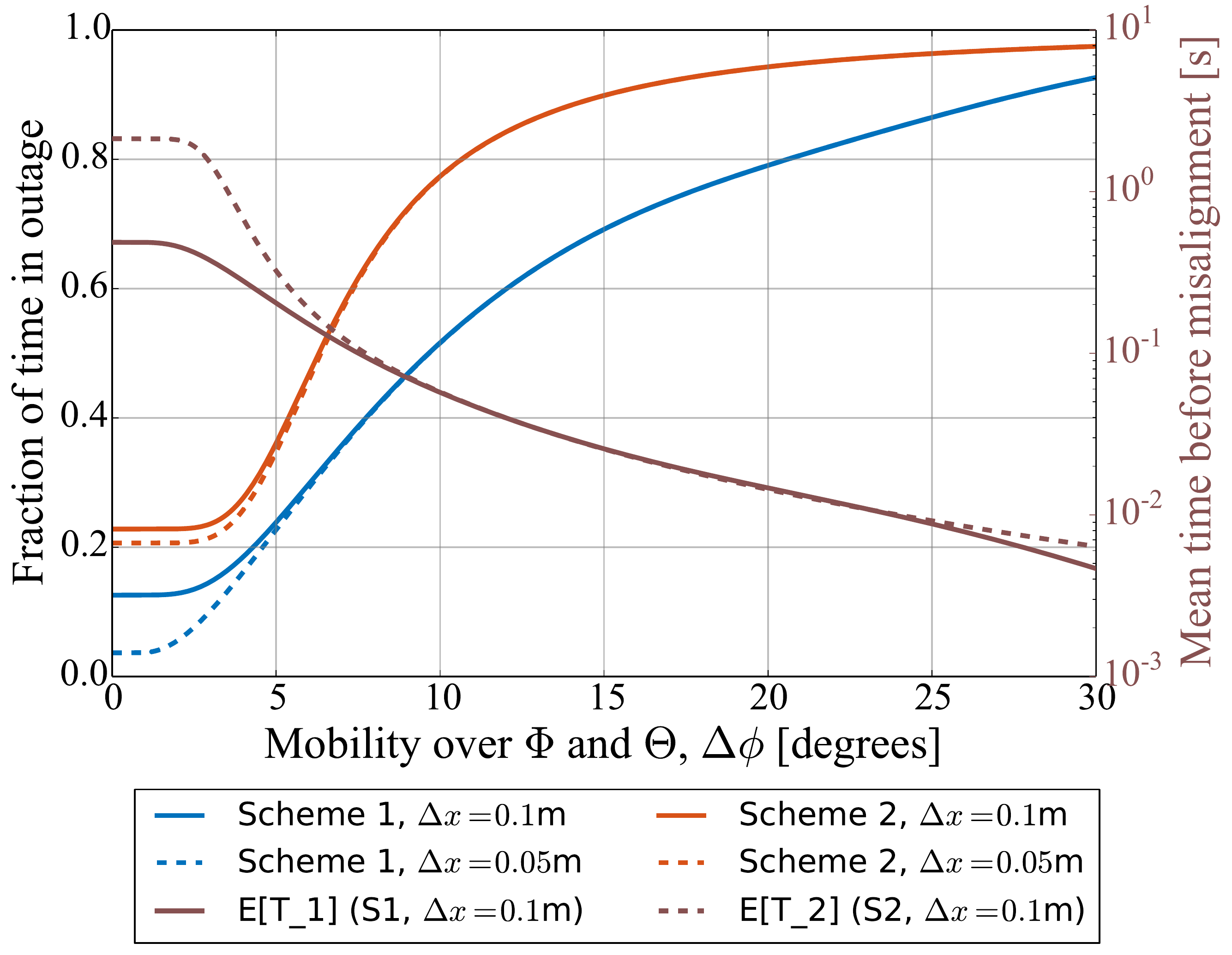}
	\label{fig:plot4}
}\hspace{-1.15mm}
\subfigure[{Mean capacity vs. $\Delta{\text{x}}$ for both schemes}]
{
	\includegraphics[width=0.31\textwidth]{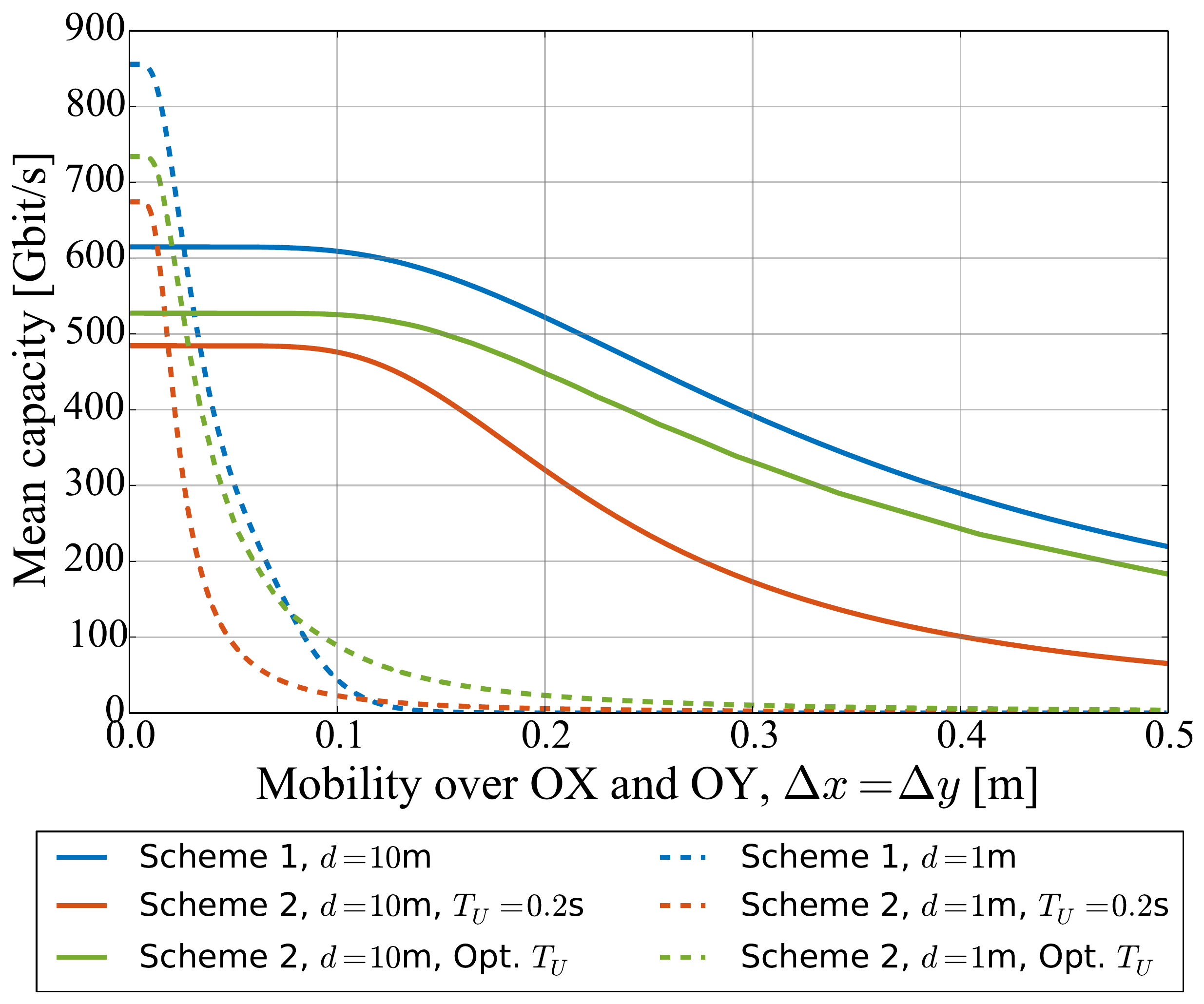}
	\label{fig:plot5}
}\hspace{-1.15mm}
\subfigure[{Mean capacity vs $d$ for both schemes}]
{
	\includegraphics[width=0.31\textwidth]{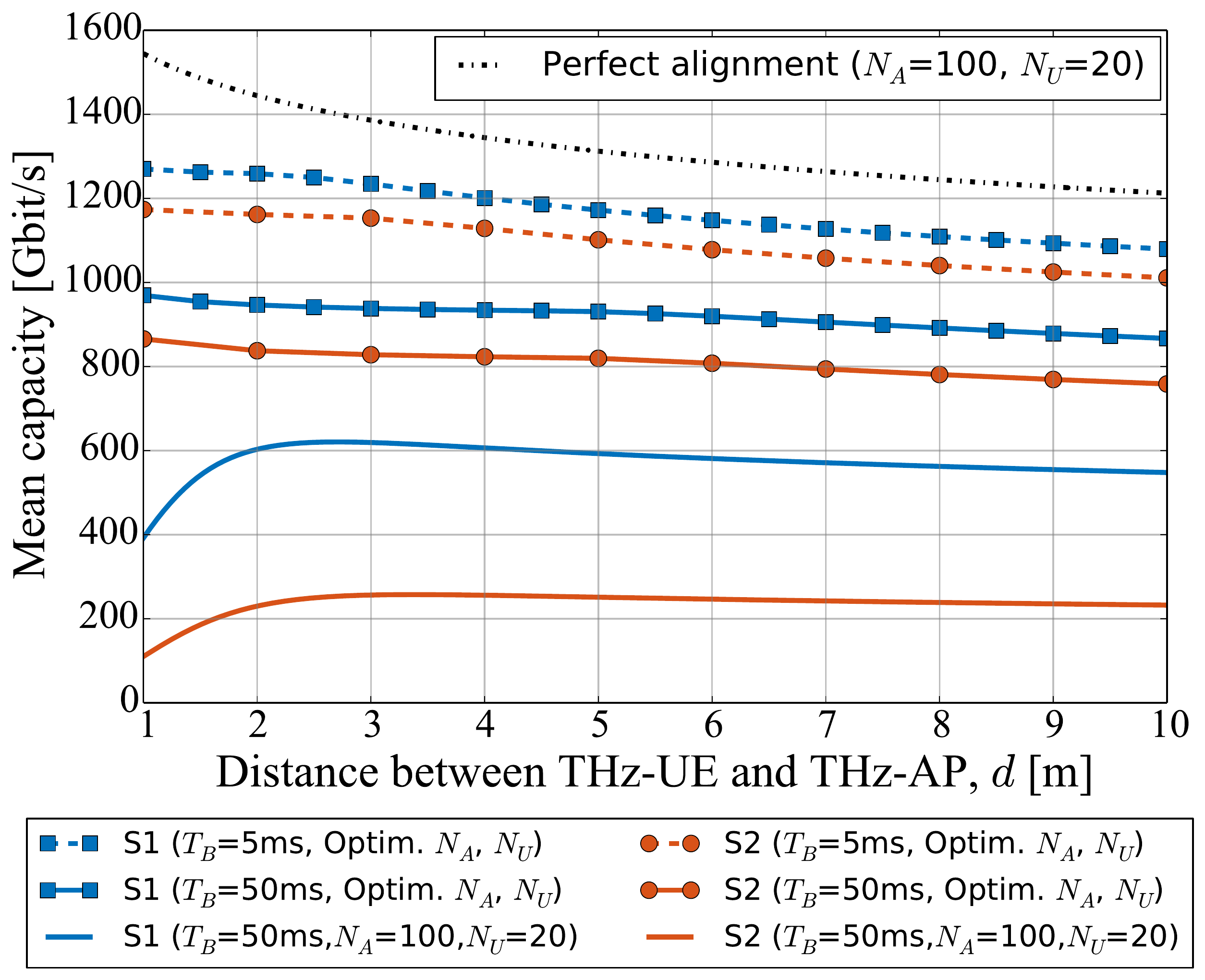}
	\label{fig:plot6}
}
\vspace{-2mm}
\caption{The impact of THz-UE micro-mobility on the outage- and capacity-centric characteristics.}
\label{fig:plots_final}
\vspace{-3mm}
\end{figure*}

\subsubsection{The effect of the beam realignment period, $T_{\text{U}}$}
We start with Fig.~\ref{fig:plot1} presenting the fraction of time in outage versus $T_{\text{U}}$ for Scheme 2. Here, we observe that too low value of $T_{\text{U}}$ results in the system spending most of its time performing beam realignment, while too rare updates ($T_{\text{U}} > 1$\,s) also lead to degraded performance. There is an ``optimal'' $T_{\text{U}}$ for each of the configurations and its typical value lies within $\approx200$-$300$\,ms and decreases with the level of micro-mobility.

\subsubsection{The effect of the antenna array size at THz-UE, $N_{\text{U}}$}
We now proceed with Fig.~\ref{fig:plot2} illustrating the impact of $N_{\text{U}}$ on the SE of the THz link when $N_{\text{A}} = 100$ and $\Delta{}x = 0.1$\,m. For Scheme 2, the ``optimal'' values of $T_{\text{U}}$ derived from Fig.~\ref{fig:plot1} are employed. In this figure, we notice that even relatively low levels of angular micro-mobility, $\Delta{}\phi = 3\degree$, lead to up to $70\%$ reduction in SE versus the theoretical limit for $N_{\text{U}} = 50$.

\textcolor{black}{One may also identify an optimization problem since greater $N_{\text{U}}$ improves the SE during the actual data transmission, but simultaneously challenges the beam alignment. Particularly, Scheme 1 achieves the highest performance with $N_{\text{U}} = 21$ for $\Delta{}\phi = 3\degree$ and with $N_{\text{U}} = 15$ for $\Delta{}\phi = 4\degree$. The greater level of angular mobility makes the lower $N_{\text{U}}$ preferable and decreases the maximum SE: $15.4$\,bit/s/Hz vs. $14.4$\,bit/s/Hz. Similar trends are observed for Scheme 2. We finally notice that Scheme 1 slightly outperforms Scheme 2 maintaining a maximum SE of approximately $1$\,bit/s/Hz higher.}

\subsubsection{The effect of the antenna array size at THz-AP, $N_{\text{A}}$}
\textcolor{black}{We continue our investigations with Fig.~\ref{fig:plot3} presenting the mean SE as a function of $N_{\text{A}}$. Similar observations to those from Fig.~\ref{fig:plot2} are made. We also notice that the chosen $N_{\text{U}}$ affects the ``optimal'' value of $N_{\text{A}}$ (and vice versa). Particularly, for Scheme 2, SE maximizes at $N_{\text{A}} = 190$ with $N_{\text{U}} = 5$ and at $N_{\text{A}} = 100$ with $N_{\text{U}} = 10$. Therefore, in order to maximize the SE $N_{\text{A}}$ and $N_{\text{U}}$ have to be optimized jointly for every set of micro-mobility characteristics. The figure also illustrates that ultra-massive antenna arrays ($N_{\text{A}} \approx 1024$, as in~\cite{akyildiz2016realizing}) are currently irrelevant for mobile systems as they invoke both low active time, $T_{\text{A}}$, and high overheads on $T_{\text{B}}$.}

\subsubsection{The effect of angular micro-mobility, $\Delta{}\phi$}
Fig.~\ref{fig:plot4} illustrates the fraction of time in outage (left axis) and the mean time to the first misalignment (right axis) as functions of $\Delta{}\phi$. $N_{\text{A}}$ is set to 100 and $N_{\text{U}}$ -- to 20. We start with the fraction of time in outage. \textcolor{black}{For both schemes, there is a regime, $\Delta{}\phi < 3\degree$, where the curves are flat as with low $\Delta{}\phi$ the system performance is mainly limited by the cartesian micro-mobility, so the minor changes in $\Delta{}\phi$ do not have any notable impact.}

\textcolor{black}{In contrast, when $\Delta{}\phi > 10\degree$, the impact of angular mobility dominates. Consequently, the curves for $\Delta{}x = 0.1$\,m and $\Delta{}x = 0.01$\,m are similar for both schemes. We also note that Scheme 1 outperforms Scheme 2 in terms of the fraction of time in outage. The opposite trend is observed for the right axis, where $E[T_{2}] > E[T_{1}]$ for all the values of $\Delta{}\phi$. The gain is the most visible for $\Delta{}\phi < 4\degree$.}

\subsubsection{The effect of Cartesian mobility, $\Delta{}x$}
We continue studying the implications of micro-mobility on THz link with Fig.~\ref{fig:plot5} presenting the mean link capacity as a function of $\Delta{}x$ when $N_{\text{A}} = 100$ and $N_{\text{U}} = 20$. Here, for $d = 10$\,m we observe similar effect as in Fig.~\ref{fig:plot4}: the small values of $\Delta{}x$ (e.g., $\Delta{}x < 0.1$\,m) do not have immediate implications on the metric, as the system is primarily limited by a relatively high $\Delta{}\phi = 6\degree$.

At the same time, for $d = 1$\,m this effect is no longer present as the beam generated by THz-AP at $1$\,m distance is less than $2$\,cm wide. Consequently, even very small drifts caused by cartesian micro-mobility lead to immediate degradation in capacity (e.g., from $850$\,Gbit/s at $\Delta{}x = 0.02$\,m to $50$\,Gbit/s at $\Delta{}x = 0.1$\,m for Scheme 1). So, it is especially important to account for cartesian micro-mobility at short distances.

\textcolor{black}{We, finally, bring the attention to comparing Scheme 2 when $T_{\text{U}} = 0.2$ with Scheme 2 when $T_{\text{U}}$ is numerically optimized for each of the $\Delta{}x$ values following the approach from Fig.~\ref{fig:plot1}. We observe the second system notably outperforming the first one for the entire range of $\Delta{}x$ and for both considered separation distances, $d$. With this comparison, we emphasize the importance of accounting for micro-mobility in THz communications and optimizing the system parameters for an envisioned pattern of micro-mobility, whenever feasible.}

\subsubsection{The effect of beam realignment time, $T_{\text{B}}$}
We continue our discussion on the necessity of system optimization accounting for the micro-mobility pattern with Fig.~\ref{fig:plot6}. This figure demonstrates the maximum achievable capacity of a THz link when all the flexible parameters -- $N_{\text{A}}$, $N_{\text{U}}$, and $T_{\text{U}}$ -- are jointly optimized following the effects in Fig.~\ref{fig:plots_optimizing}.

{
\color{black}
Two major observations are made from this figure. First, the curves for $T_{\text{B}} = 50$\,ms are $\approx$$350$\,Gbit/s lower that the ones for $T_{\text{B}} = 5$\,ms. So, the design of reliable and efficient procedure for beam realignment is of crucial importance for mobile THz communications. Secondly, there is a $0.6$--$1.1$\,Tbit/s difference between the theoretical curve and those for practical $T_{\text{B}} = 50$\,ms (corresponds to $N_{\text{A}} = 100$, $N_{\text{U}} = 20$, and $\delta$ = $5$\,ms). Thus, it is important to account for micro-mobility in performance evaluation of mobile THz systems. Ignoring this factor results in severe overestimation of the achievable data rates.

Finally, we compare the performance of Schemes 1 and 2 with $T_{\text{B}} = 50$\,ms and optimized $N_{\text{A}}$, $N_{\text{U}}$ when $T_{\text{B}} = 50$\,ms, $N_{\text{A}}=100$, and $N_{\text{U}}=20$. As seen from Fig.~\ref{fig:plot6}, the former one provides up to \emph{half of  a Tbit/s} greater capacity. The gain is especially visible at $d < 2$\,m, where cartesian micro-mobility results in worse SE for smaller distances despite the lower pathloss. The latter observation confirms the need to account for THz-UE micro-mobility not only in performance studies, but also in the design of the transceivers for prospective THz mobile communications.
}

\vspace{-2mm}
\section{Conclusions}\label{sect:05}\label{sec:conclusions}

\balance
In this paper, we proposed a mathematical framework to estimate the performance of THz communications in the presence of both Cartesian and angular micro-mobility of the THz-UE. Our study revealed that: (i) the micro-mobility of THz-UE may decrease the capacity of the THz link by hundreds of Gbit/s; (ii) the Cartesian micro-mobility is more harmful at shorter distances, when $d$ is around few meters, while the impact of angular one is more visible at longer ones; (iii) on-demand beam alignment procedure leads to $10$--$15$\% higher spectral efficiency than periodic beam alignment. The framework can be tuned for other micro-mobility patterns (autonomous driving, UAV networks, etc.) by replacing $f_{T_{\text{A}}}(t)$ in (\ref{eqn:overallpdf}) without any modifications in the further~analysis. \textcolor{black}{It can also be further extended to analyze the enhanced beam management procedures, e.g., those reported in~\cite{revision_1,revision_2}. We plan to rely on this framework in the future when developing our end-to-end THz communication system with micro-mobility.}

\textcolor{black}{There are several important practical outcomes of our study. First of all, it is shown that the periodic beam realignment procedure (Scheme 2) must be performed every $100$\,ms--$300$\,ms depending on the system parameters to maximize the link capacity. It is also revealed that optimizing $T_{\text{U}}$ using our framework allows for up to $200\%$ lower fraction of time in an outage. Secondly, it is illustrated that a too high number of antenna elements on either THz-AP or THz-UE decreases the link performance. In contrast, when one selects the number of antenna elements using the contributed framework, it becomes possible to improve the capacity of the THz link by as high as $500$\,Gbit/s. Finally, the framework also enables determining the specific numerical requirements for the beam realignment time and THz beam steering delay, thus providing the necessary insights for the design of the prospective hardware for future mobile THz systems.}

\vspace{-3mm}
\bibliographystyle{ieeetr}
\bibliography{imperfect_short}

\end{document}